\documentclass{emulateapj}
\usepackage{apjfonts}
\usepackage{epsf}
\begin{document}

\slugcomment{}
\shortauthors{J. M. Miller et al.}
\shorttitle{A NuSTAR Observation of GRS 1739$-$278}

\title{New Constraints on the Black Hole Low/Hard State Inner
  Accretion Flow with NuSTAR}

\author{J.~M.~Miller\altaffilmark{1}, 
J.~A.~Tomsick\altaffilmark{2},
M. Bachetti\altaffilmark{3},\altaffilmark{4},
D.~Wilkins\altaffilmark{5}, 
S.~E.~Boggs\altaffilmark{2},
F.~E.~Christensen\altaffilmark{6},
W.~W.~Craig\altaffilmark{7},\altaffilmark{8},
A. C. Fabian\altaffilmark{9},
B.~W.~Grefenstette\altaffilmark{10},
C.~J.~Hailey\altaffilmark{8}, 
F.~A.~Harrison\altaffilmark{10}, 
E.~Kara\altaffilmark{9}, 
A.~L.~King\altaffilmark{1}, 
D.~K~Stern\altaffilmark{11},
W.~W.~Zhang\altaffilmark{12}}

\altaffiltext{1}{Department of Astronomy, University of Michigan, 500
Church Street, Ann Arbor, MI 48109-1042, USA, jonmm@umich.edu}

\altaffiltext{2}{Space Sciences Laboratory, University of California,
  Berkeley, CA 94720, USA}

\altaffiltext{3}{Universite de Toulouse, UPS-OMP, IRAP, F-31100 Toulouse, France}

\altaffiltext{4}{CNRS, Intitut de Recherche in Astrophysique et
  Planetologie, 9 Av. Colonel Roche, BP 44346, F-31028 Toulouse, Cedex
  4, France}

\altaffiltext{5}{Department of Astronomy \& Physics, Saint Mary's
  University, Halifax, NS. B3H 3C3, Canada}

\altaffiltext{6}{Danish Technical University, DK-2800, Lyngby, Denmark}

\altaffiltext{7}{Lawrence Livermore National Laboratory, Livermore CA, USA}

\altaffiltext{8}{Columbia University, New York, NY 10027, USA}

\altaffiltext{9}{Institute of Astronomy, University of Cambridge,
  Madingley Road, Cambridge CB3 OHA, UK}

\altaffiltext{10}{Cahill Center for Astronomy and Astrophysics,
  California Institute of Technology, Pasadena CA 91125, USA}

\altaffiltext{11}{Jet Propulsion Laboratory, California Institute of Technology, 4800 Oak Grove Drive, Pasadena CA, 91109, USA}

\altaffiltext{12}{NASA Goddard Space Flight Center, Greenbelt MD, 20771, USA}


\label{firstpage}

\begin{abstract}
We report on an observation of the Galactic black hole candidate GRS
1739$-$278 during its 2014 outburst, obtained with {\it NuSTAR}.  The
source was captured at the peak of a rising ``low/hard'' state, at a
flux of $\sim0.3$~Crab.  A broad, skewed iron line and disk reflection
spectrum are revealed.  Fits to the sensitive {\it NuSTAR} spectra
with a number of relativistically blurred disk reflection models yield
strong geometrical constraints on the disk and hard X-ray ``corona''.
Two models that explicitly assume a ``lamppost'' corona find its base
to have a vertical height above the black hole of $h = 5^{+7}_{-2}~
GM/c^{2}$ and $h = 18\pm 4~ GM/c^{2}$ (90\% confidence errors); models
that do not assume a ``lamppost'' return emissivity profiles that are
broadly consistent with coronae of this size.  Given that X-ray
microlensing studies of quasars and reverberation lags in Seyferts
find similarly compact coronae, observations may now signal that
compact coronae are fundamental across the black hole mass scale.  All
of the models fit to GRS 1739$-$278 find that the accretion disk
extends very close to the black hole -- the least stringent constraint
is $r_{in} = 5^{+3}_{-4}~ GM/c^{2}$.  Only two of the models deliver
meaningful spin constraints, but $a = 0.8\pm 0.2$ is consistent with
all of the fits.  Overall, the data provide especially compelling
evidence of an association between compact hard X-ray coronae and the
base of relativistic radio jets in black holes.
\end{abstract}

\section{Introduction}
The phenomenon of relativistic radio jets in active galactic nuclei
(AGN) has long been appreciated, but the ability of black hole jets to
shape host galaxies and even clusters of galaxies has only become
clear recently (Fabian 2012).  Accretion onto distant, low-luminosity,
jet-producing active galactic nuclei (AGN) is difficult to probe,
however, owing to low X-ray flux levels.  In contrast, Galactic
stellar-mass black holes can provide very sensitive observations.  In
their low-$\dot{m}$, spectrally-hard state (``low/hard'' state),
stellar-mass black holes launch compact, steady jets like those
observed in some AGN.  If accretion flows are self-similar for given
Eddington ratios, studies of Galactic black holes in the low/hard
state may be the key to understanding jet production in AGN.

GRS 1739$-$278 was discovered with the SIGMA telescope on board {\it
  Granat} (Paul et al.\ 1996, Vargas et al. 1997).  The X-ray spectral
and timing characteristics of GRS 1739$-$278 strongly suggest that the
source harbors a black hole primary.  In particular, a very strong
5~Hz quasi-periodic oscillation (QPO) was detected in GRS 1739$-$278,
which is typical of black holes when they enter a ``very high'' state
(Borozdin \& Trodolyubov 2000; Wijnands et al. 2001).  Moreover, the
source was detected in radio during its 1996 outburst (Hjellming et
al.\ 1996), and strong radio emission in transients is typical of
black holes; radio emission in neutron stars is notably weaker
(Migliari \& Fender 2006).  The Galactic center location of GRS
1739$-$278, and a measured extinction of $A_V = 14\pm2$ strongly
suggest that the source is located at a distance of 6--8.5~kpc
(Dennerl \& Greiner 1996; Greiner, Dennerl, \& Predehl 1996).

Following an extended quiescent period, GRS 1739$-$278 was detected in
outburst with the {\it Swift} Burst Alert Telescope (15--50~keV)
on 2014 March 9 (Krimm et al.\ 2014).  The source continued to
brighten steadily, reaching a flux of approximately 0.065~c/s during
its rise phase (the Crab gives a count rate of $\simeq$0.22~c/s).  The source
was also noticed during monitoring observations of the Galactic center
made with {\it INTEGRAL}; a hard ($\Gamma = 1.4\pm 0.2$) cut-off
power-law spectrum was detected out to 200~keV (Filippova et
al.\ 2014).  

A prior {\it NuSTAR} (Harrison et al.\ 2013) observation of GRS
1915$+$105 in a ``low/hard'' (or, ``plateau'') state revealed a disk
reflection spectrum in unprecedented detail (Miller et al.\ 2013).
The sensitivity of that spectrum permitted a precise black hole spin
measurement, and indicated a small, centrally concentrated corona
consistent with a ``lamppost'' geometry.  {\it NuSTAR} has also
measured black hole spin values in Cygnus X-1 and 4U 1630$-$472
(Tomsick et al.\ 2014; King et al.\ 2014).  Motivated by these
results, we triggered a {\it NuSTAR} observation of GRS 1739$-$278 to
better explore black hole accretion flows in the jet mode.

\section{Data Reduction}
The {\it NuSTAR} observation of GRS 1739$-$278 started on 2014 March
26 at 16:06:07 UT.  Event files and light curves were generated using
the tools and routines in HEASOFT version 6.15.1, particularly
NuSTARDAS version 1.3.1, and CALDB version 20131223.  After filtering,
the FPMA and FPMB detectors obtained net exposure times of 29.7~ks and
30.4~ks, respectively.

Circular extraction regions with a radius of 120 arcseconds, centered
on the source position, were used to extact source counts.  The very
bright nature of the source likely provides no region on the focal
plane that is truly free of source counts; regions of equivalent size
but far from the source were extracted as background.

Prior to spectral fitting within XSPEC version 12.8.1g (Arnaud 1996),
the spectra were grouped to require a signal-to-noise ratio of at
least 30 in all bins.  This procedure exceeds the minimum quality
necessary for $\chi^{2}$ statistical analysis to be valid (Cash 1979),
but helps to build signal at high energy.  All of the uncertainties quoted in
this work are 90\% confidence errors.

\section{Analysis and Results}
Initial fits revealed some potential minor deficiencies in the
response characterization.  The highest energy bins lie slightly above
plausible models; checks of the background show that these deviations
are not likely to have a related origin.  The problem was mitigated by
ignoring the highest energy (grouped) bin.  In our fits, then, the
highest energy bins only extend up to $\sim$70~keV, rather than the
nominal 79~keV.  Similarly, small deviations were found at the lowest
energies covered by FPMA and FPMB.  In particular, the best fit
line-of-sight column density in the FPMA spectrum is ${N}_{\rm H} \sim
2.3\times 10^{22}~ {\rm cm}^{-2}$; however, values lower by
0.3--0.4$\times 10^{22}~ {\rm cm}^{-2}$ are preferred in the FPMB
spectrum.  Different source and background regions did not
significantly alter this disparity.  It is plausible that the
difference results from small calibration errors.  We therefore
allowed the column density ${N}_{\rm H}$ to vary between the spectra,
and for simplicity we only report the values measured for the FPMA
detector.  In view of these minor issues, we allowed the power-law
index to float between the FPMA and FPMB detectors in all fits;
however, any differences were found to be negligible, with $\Delta
\Gamma \leq 0.02$.  Last, we allowed an overall constant to float
between the FPMA and FPMB to account for differences in their absolute
flux calibrations.  


Figure 1 shows the spectra of GRS 1739$-$278, fit with a simple
cut-off power-law model, appropriate for black holes in the low/hard
state (see Table 1).  There is no evidence of thermal emission from an
accretion disk; this likely owes to the low disk temperatures
typically measured in the low/hard state (e.g. $kT = 0.2-0.3$~keV;
Miller et al.\ 2006a,b; Reis et al.\ 2010; Reynolds \& Miller 2013),
and a high level of line-of-sight obscuration.  

Disk reflection features are clearly revealed in the data/model ratio
in Figure 1.  The rest of our analysis therefore focused on extracting
information from the disk reflection spectrum.  {\it Reflionx} is an
established model that is suited to the high ionization levels
expected in X-ray binaries (Ross \& Fabian 2005).  In this work, an
updated version of {\it reflionx} is used that includes a spectral
cut-off in the illuminating power-law as a fitting parameter (see,
e.g. Miller et al.\ 2013).  Fits were also made using the {\it
  xillver} reflection model (Garcia et al.\ 2013, 2014), specifically
{\it xillver-a-Ec}, which again includes an exponential cut-off.  {\it Xillver} is
relatively new; it features higher spectral resolution, newer atomic
data, and includes angle dependencies.  In contrast, {\it reflionx} is
an ``angle-averaged'' model.

Reflection models are calculated in the fluid frame of the accretion
disk, and a blurring function is required to tranlsate from the fluid
frame to the observed frame.  These convolution models encode not only the
relativistic Doppler shifts and gravitational red shifts expected
close to black holes (ultimately allowing for spin measurements), but
also the geometry of the hard X-ray corona.  Whereas energy shifts
encode the potential, the emissivity law captures the geometrical
dependencies (see, e.g., Wilkins \& Fabian
2012, Dauser et al.\ 2013, 2014).  

In this work, both {\it reflionx} and {\it xillver} were convolved with
{\it relconv} to shift frames.  The action of
{\it relconv} on {\it xillver} is captured by the combined model
{\it relxill} (Dauser et al.\ 2013).  {\it Xillver} was also used in
conjunction with two convolution models that assume a ``lamppost''
geometry (hard X-ray emission from directly above the black hole, along its
spin axis).  The associated emissivity law is hard-wired into
these functions.  Fits that assume a static ``lamppost'' corona were
made using {\it relxilllp}, and fits exploring an outflowing
``lamppost'' with non-negligible vertical extent were made by applying
the function {\it relconv\_lp\_ext} to {\it xillver}.

The {\it relxill} and {\it relconv*reflionx} models require the
addition of an external cut-off power-law.  In both cases, the
power-law index and cut-off energy were linked between the reflection
model and the external continuum component.  These models do not
assume a geometry for the corona, and therefore include an emissivity
function encoded as a broken power-law, giving parameters $q_{in}$,
$q_{out}$, and $r_{break}$.  Guided by the many different scenarios
considered by Wilkins \& Fabian (2012), we required $3 \leq q_{in}
\leq 10$, $0 \leq q_{out} \leq 3$, and $2 \leq c^{2}r_{break}/GM \leq
6$.  In contrast, the fits made using {\it relconv\_lp\_ext*xillver} and
{\it relxilllp} do not require emissivity indices, and instead constrain
the height of the coronal base $h$ in units of $GM/c^{2}$.  The
{\it relconv\_lp\_ext*xillver} model is the most sophisticated of all;
it essentially assumes that the corona is the base of a relativistic
jet.  It allows the data to constrain the vertical extent of the
corona (not just the height), and also allows the data to constraint
the velocity of the corona at its base and at its top.  Preliminary
tests suggested that the data could not constrain all of these
parameters, so we only considered a corona with a single velocity
($v_{\rm corona}/c$ in Table 1) and a fixed extent of $10~GM/c^{2}$.

Table 1 lists the results of fitting all four models, and Figure 2
shows the corresponding unfolded spectra and data/model ratios.  The
models all measure similar power-law indices and values for the
cut-off energy.  The values of $N_{\rm H}$ are also broadly similar.
This level of consistency likely signals that the continuum is
well-measured, and not strongly model-dependent.  Different reflection models
give slightly different values for the ionization and abundances in
the accretion disk.  All three models based on
{\it xillver} measure ionization values consistent with ${\rm log}(\xi)
= 3.12$, whereas the fit utilizing {\it reflionx} measures ${\rm
  log}(\xi) = 3.45(5)$.  The implied Fe abundances also differ between
the models using {\it xillver} and {\it relfionx}.  

All of the fits strongly suggest that the inner accretion disk in GRS
1739$-$278 remained close to the ISCO in this bright phase of the
``low/hard'' state.  The least stringent constraint comes from fits with
{\it relxillp}, which nominally allow for $r_{in} = 8~GM/c^{2}$ at the
90\% level of confidence.  This is only slightly larger than the ISCO
for prograde accretion onto an $a = 0$ black hole.  Figure 3 shows how
the fit statistic varies for a larger range of possible inner disk
radii.  The best-fit model rules out a value of $r_{in} \sim
12~GM/c^{2}$ at the $5\sigma$ level of confidence.

The models that explicitly assume a ``lamppost'' geometry measure
small heights for the base of the corona: {\it relconv\_lp\_ext*xillver}
gives $h = 5^{+7}_{-2}~GM/c^{2}$ (recall our model assumed a vertical
extent of $10~ GM/c^{2}$ above this base), and {\it relxilllp} gives $h
= 18\pm 4~ GM/c^{2}$.  A very steep inner emissivity and flatter outer
emissivity are commensurate with a ``lamppost'' emitter above a spinning
black hole (Wilkins \& Fabian 2012; Dauser et al.\ 2013, 2014).  The
emissivity profiles measured from the other models in Table 1 are
consistent with $q_{in} = 8$ and $q_{out} = 2$, indirectly indicating
a lamppost-like geometry with light-bending close to a spinning black
hole.  These same models also provide strong direct constraints on the
spin of the black hole, measuring high values consistent with $a =
0.8\pm 0.2$.

The models that explicitly assume a lamppost geometry do not provide
meaningful constraints on the spin of the black hole in GRS 1739$-$278
(see Table 1 and Figure 3), though they certainly allow for high spin
values.  This may indicate potential degeneracies between
geometric constraints and spin constraints in the ``low/hard'' state
(Dauser et al.\ 2013, 2014; Fabian et al.\ 2014).  In this sense, the
spin measurements quoted above should be regarded with a degree of
caution.  Note, however, that although the lamppost models do not
place strong constraints on the black hole spin, they still place
strong constraints on the inner radius of the disk, ruling out large
inner disk radii.

Both {\it relxilllp} and {\it relxill} indicate low values for the
reflection fraction, much less than unity ($f_{\rm reflection} =
reflected/incident$).  This may be partly artificial: the models
calculate the reflection fraction in the 20--40~keV band (Dauser et
al.\ 2013, 2014).  Extrapolating the fit with {\it reflionx} over the
0.001--1000~keV band, a reflection fraction of $\sim1.2$ is obtained.
If the low fraction is physical, it could be explained via a truncated
accretion disk that does not subtend the expected solid angle, or a
corona that has beamed part of its emission away from the disk, or a
stratified disk atmosphere that diminishes the overall reflection
spectrum (Ballantyne \& Fabian 2001, Nayakshin \& Kallman 2001).  Given
that all of the models strongly require small inner disk radii and
small coronae, the low reflection fraction may indicate an outflowing
corona, as predicted if the corona is the base of a relativistic jet
(Markoff, Nowak, \& Wilms 2005; also see Miller et al.\ 2006, 2013).
Indeed, the best-fit model in Table 1 is {\it
  relconv\_lp\_ext*xillver}, which assumes an outflowing corona (note,
however, that the data are not able to constrain the velocity of the
outflow).

Finally, we attempted to explicitly examine whether or not the data
can reject a radially extended corona.  Wilkins \& Fabian (2012) give
different emissivity prescriptions, corresponding to different coronal
geometries.  Their work captures the emissivity in terms of two break
radii, whereas a model like {\it relconv} only has one break.  Based
on the best fit to 1H~0707$-$495 detailed in Wilkins \& Fabian (2012),
fake spectra were generated assuming $q_{in} = 7.8$, $r_{break} =
5~GM/c^{2}$, $q_{mid} = 0$, and $q_{out} = 3.3$, but varying the outer
break radius ($J \propto r^{-q}$).  The fake spectra were then fit
with an emissivity profile that has only one break, in order to
establish how a simpler emissivity prescription might encode a
radially extended corona.

A ``true'' coronal radial extent of $50~GM/c^{2}$ is encoded as having
$q_{in} = 5.09$, $q_{out} = 1.73$, and $r_{break} = 5.09~GM/c^{2}$
when only a single break is used.  Fits to GRS 1739$-$278 with the
{\it relxill} model give $\chi^{2} = 1353.0/1107$ for this emissivity
prescription (all other parameters were free to vary).  The best-fit
``lamppost'' model is preferred over this example of a radially-extended
corona at the $4.2\sigma$ level of confidence.  However, the simplest
geometry with a truncated disk is one wherein the corona lies entirely
within the truncated disk.  This corresponds to an ADAF-like geometry
in which the inner disk has become too hot to be cool and thin.  If
the inner radius is fixed to the (encoded) break radius of
$5.09~GM/c^{2}$, the fit becomes even worse: $\chi^{2}/\nu =
1400.5/1108$.  A ``lamppost'' geometry is preferred to this
prescription at the $7.1\sigma$ level of confidence.

\section{Discussion and Conclusions}
We have analyzed spectra of the transient and recurrent black hole
candidate GRS 1739$-$278, obtained near the peak of a rising
``low/hard'' state with {\it NuSTAR}.  The extraordinary sensitivity
and broad energy range of the spectra reveal a reflection spectrum in
detail, and strong geometric constraints are derived.  The data
strongly signal reflection from an accretion disk that remains close
to the black hole, irradiated by a compact and potentially outflowing
corona at a moderate height above the black hole.  Compared to ``very
high'' or ``intermediate'' states, then, the peak of the ``low/hard''
state may be characterized by weaker illumination of the disk (perhaps
due to beaming of coronal/jet emission), and a larger coronal scale
height.  In this section, we review these results and discuss their
implications.

The results listed in Table 1 and shown in Figure 3 indicate
that the disk in GRS 1739$-$278 remained close to plausible ISCO
values.  Sensitive {\it NuSTAR} spectra of the persistent stellar-mass
black hole GRS 1915$+$105 in a ``low/hard'' state also required a disk
extending to the ISCO around a rapidly spinning black hole (Miller et
al.\ 2013).  Moreover, a series of 20 {\it Suzaku} observations of the
archetypical black hole Cygnus X-1 in the ``low/hard'' state revealed
a blurred disk reflection spectrum that requires a disk that extends
close to the ISCO (Miller et al.\ 2012); Fabian et al.\ (2012) used
these data to measure a spin of $a = 0.97^{+0.01}_{-0.02}$ and an
inner radius of $r_{in} \leq 2~GM/c^{2}$.

Esin, McClintock, \& Narayan (1997) predicted that disks may truncate
at $\dot{m} \leq 0.08$, corresponding to $L/L_{Edd} \leq 0.008$
(assuming an efficiency of $\eta = 0.1$).  The best-fit model for GRS
1739$-$278 in Table 1 implies $L = 1.0\times 10^{38}~ {\rm erg}~ {\rm
  s}^{-1}$ (0.5--100.0~keV) or $L/L_{Edd} \simeq 0.08$ for a black
hole of $10~ M_{\odot}$.  Indeed, all of the results cited above were
obtained in luminous phases of the ``low/hard'' state, above the
predicted truncation luminosity.  Our results simply indicate, then,
that state transitions -- and jet production -- are driven primarily
by coronal changes, rather than large or rapid changes in the disk.
This may also be supported by relatively smooth variations in disk
temperatures across state transitions (e.g. Rykoff et al.\ 2007; also
see Gierlinski et al.\ 2009).  This only requires that $\dot{M}$
varies while $R$ remains relatively constant, since the disk
temperature and viscious dissipation depend on {\it both} $\dot{M}$
{\it and} $R$.  Deeper into the ``low/hard'' state, $\dot{M}$ and $R$
might decrease together.  Positive evidence of disk truncation at
$L/L_{Edd} \simeq 0.001$ was indicated by a narrow Fe K line in a {\it
  Suzaku} spectrum of GX~339$-$4 (Tomsick et al.\ 2009).

The reflection fits also provide important indications for a very
compact, and potentially outflowing corona that is likely the the base
of a relativistic jet.  The small coronal heights that are measured
with two models, the steep emissivity profiles measured with two
others, small measured reflection fractions (potentially), and the
fact that the best overall model assumes an outflowing corona, all
point in this direction.  X-ray reverberation from the inner disk in
Seyfert AGN strongly indicates short distance scales, and small
coronal heights (e.g. Zoghbi et al.\ 2010, De Marco et al.\ 2013).
And, groundbreaking X-ray microlensing measurements in quasars also
constrain the hard X-ray corona to have a size of only
$\simeq$10--20~$GM/c^{2}$ (e.g. Morgan et al.\ 2010, Blackburne et
al.\ 2014).  It is possible, then, that compact coronae are a central,
mass-independent feature of black hole accretion and jet production.

Coronal energetics also point toward an association with the base of a
jet.  Merloni \& Fabian (2001) considered the coronal scales necessary
to match implied X-ray luminosities, assuming that X-ray coronae are
powered by pure thermal Comptonization.  Using the cut-off energy
values given in Table 1, and further assuming $\tau \simeq 0.1$, the
corona in GRS 1739$-$278 would need to be $10^{4}~ GM/c^{2}$ in radial
extent to generate the implied luminosity.  Even $\tau \simeq 1$ gives
$10^{3}~GM/c^{2}$.  The {\it NuSTAR} spectra do not allow for a hole
of this radius within the disk, nor a spheroid of such a height.
Based on such contradictions, and the small length scales implied by X-ray
variability, Merloni \& Fabian (2001) found that coronae must be
primarily non-thermal, and likely magnetic.  A broad range of models
predict that magnetic fields play a role in jet production.  Very
compact coronae may hint at close interactions with black holes, as
per Blandford \& Znajek (1977).

JMM thanks Javier Garcia and Thomas Dauser for helpful conversations.
This work was supported under NASA Contract No. NNG08FD60C, and made
use of data from the NuSTAR mission, a project led by the California
Institute of Technology, managed by the Jet Propulsion Laboratory, and
funded by NASA.

\clearpage

\begin{table}[t]
\caption{Spectral fits to GRS 1739$-$278}
\begin{footnotesize}
\begin{center}
\begin{tabular}{llllll}
\tableline
  ~                            & relconv\_lp\_ext & relxill\_lp & relxill    & relconv          & cut-off-  \\
  ~                            & $\times$xillver  &     ~       &    ~       & $\times$reflionx & power-law          \\
\tableline
$N_{\rm H}$ ($10^{22}~{\rm cm}^{-2}$) &     2.18(6)    &   2.35(6)  &   1.94(6)  &     1.95(5)      &           4.6             \\
           ~                     &         ~        &    ~       &    ~      &         ~         &               ~                     \\     
$\Gamma$                         &       1.46(1)    &  1.47(1)   &  1.44(1)  &     1.44(1)          &           1.65            \\
           ~                     &         ~        &       ~    &    ~      &         ~         &            ~                     \\     
$E_{cut}$ (keV)                   &      27.5(5)     &  28.0(3)   &   27.2(3) &    31.3(3)         &           36           \\
           ~                     &         ~        &       ~    &    ~      &         ~         &            ~                     \\     
$K_{power-law}$                     &      0.75(3)     &    --      &    --     &     0.658(5)      &          1.3           \\
           ~                     &         ~        &       ~    &    ~       &         ~         &         ~                     \\     
$q_{in}$                          &       --         &    --      &    8.5(8)     &    $8(1)$             &          --          \\ 
           ~                     &         ~        &       ~    &    ~     &         ~         &             ~                     \\     
$q_{out}$                         &       --         &    --      &   2.0(1)     &     1.9(1)            &           --           \\
           ~                     &         ~        &       ~    &    ~     &         ~         &              ~                     \\     
$r_{break}$ ($GM/c^{2}$)             &       --         &     --    &  $6_{-0.1}$  &     $6_{-0.1}$       &         --           \\
           ~                     &         ~        &       ~    &    ~     &         ~         &             ~                     \\    
$a$ ($cJ/GM^{2}$)                 &      $0(1)$     &    $0(1)$   &  0.8(2)   &    0.94(2)            &            --          \\
           ~                     &         ~        &       ~     &     ~     &         ~         &              ~                     \\     
$r_{in}$ ($GM/c^{2}$)              &   $2^{+5}_{-1}$    &  $5^{+3}_{-4}$  & 2.4(1)      &    $1.0^{+0.1}$     &              --          \\
           ~                     &         ~        &       ~       &    ~     &         ~         &               ~                     \\     
$h_{\rm corona}$ ($GM/c^{2}$)        &    $5^{+7}_{-2}$   &   18(4)     &    --     &    --          &                --          \\
           ~                     &         ~        &       ~      &    ~     &         ~         &             ~                     \\     
$v_{\rm corona}/c$                & $0.98^{+0.02}_{-0.98}$ &   --     &   --      &       --        &             --           \\
           ~                     &         ~        &       ~      &    ~      &         ~         &              ~                     \\     
$i$ (degrees)                    &      33(1)       &    32.5(5)  &  43.2(5)  &   24(3)           &         --           \\
           ~                     &         ~        &     ~         &    ~     &         ~         &             ~                     \\     
log($\xi$)                       &      3.13(2)     & 3.12(2)    &  3.14(2)  &  3.45(5)             &            --           \\
           ~                     &         ~        &       ~       &    ~     &         ~         &            ~                     \\     
$A_{\rm Fe}$                       &      1.8(2)       &   1.5(2)      &   1.9(2)   &   0.61(3)       &             --          \\
           ~                     &         ~        &       ~       &    ~     &         ~         &             ~                     \\     
$f_{\rm reflection}$                 &      --        &  0.174(4)      &  0.18(2)     &   --          &               --          \\
           ~                     &         ~        &       ~         &    ~     &         ~         &            ~                     \\     
$K_{reflection}$                    & $1.07(7)\times10^{-6}$ &  0.74(2)     &   0.68(2)   & $5.09(3) \times10^{-6}$  &                --          \\
           ~                     &         ~        &       ~         &    ~     &         ~         &             ~                     \\     

$\chi^{2}/\nu$                &  1327.5/1105    &   1330.3/1106       & 1338.3/1105 &  1339.3/1104  &        7869.7/1114            \\
\tableline
\end{tabular}
\vspace*{\baselineskip}~\\ \end{center} 
\tablecomments{The results of a simple continuum fit (a cut-off
  power-law), and with four different relativistically--blurred disk
  reflection models are given in the table above (see the text for
  details).  All quoted errors are 90\% confidence limits.  The
  normalization of different components is given in columns labeled
  with ``$K$'', $A_{\rm Fe}$ is the iron abundance relative to solar,
  and $f_{\rm refl}$ is the reflection fraction.}
\vspace{-1.0\baselineskip}
\end{footnotesize}
\end{table}
\medskip

\clearpage

\begin{figure}
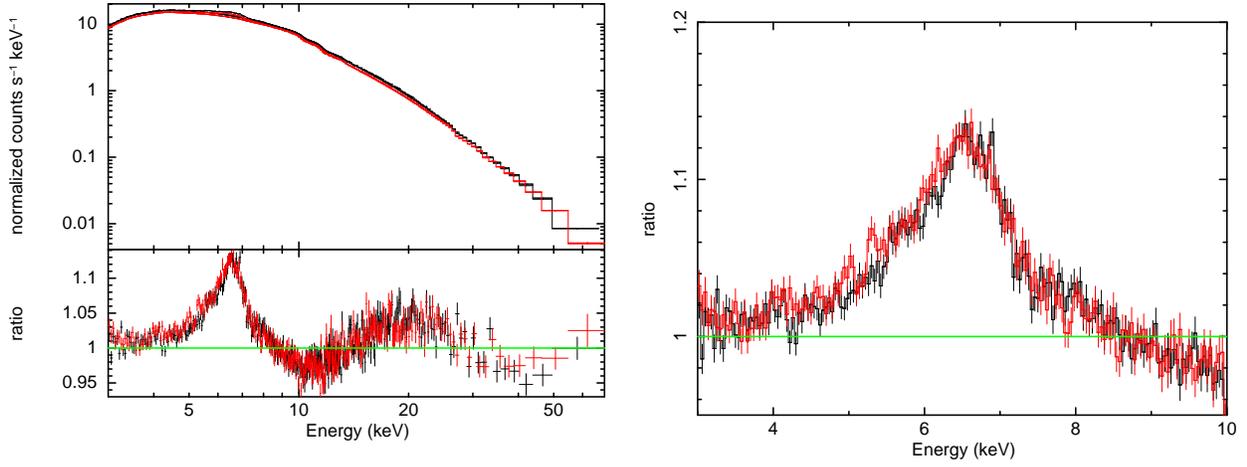

\includegraphics[scale=0.33,angle=-90]{f1a.ps}
\includegraphics[scale=0.33,angle=-90]{f1b.ps}
\figcaption[t]{\footnotesize Simple cut-off power-law fits to the continuum emission in the {\it
    NuSTAR} spectrum of GRS 1739$-$278 (FPMA in black; FPMB in red).
  The lefthand panel shows the data and ratio that result when the Fe
  K band (4--8~keV) is ignored while fitting the continuum.  A skewed
  Fe K line and reflection ``hump'' are clearly visible in the
  spectra.  The righthand panel shows the data/model ratio from the
  same fit, on a small linear scale in the Fe K band.  The breadth and
  characteristic shape of the relativistic line are clear.  The
  $N_{\rm H}$ and $E_{\rm cut}$ values in Table 1 were assumed in fitting the
  cut-off power-law continuum.}
\end{figure}
\medskip

\begin{figure}
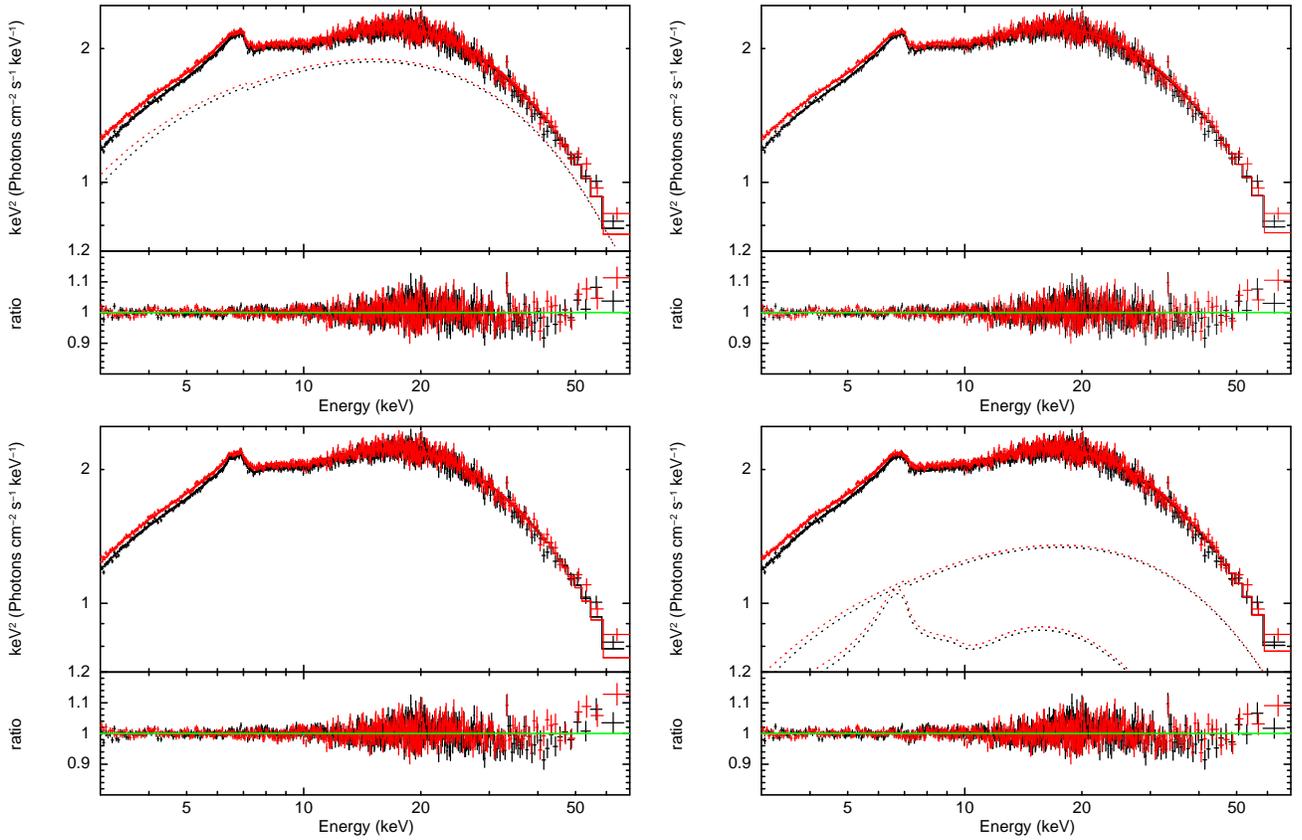

\includegraphics[scale=0.33,angle=-90]{f2a.ps}
\includegraphics[scale=0.33,angle=-90]{f2b.ps}\\
\includegraphics[scale=0.33,angle=-90]{f2c.ps}
\includegraphics[scale=0.33,angle=-90]{f2d.ps}
\figcaption[t]{\footnotesize The {\it NuSTAR} spectra of GRS
  1739$-$278 are shown, fit with the relativistically blurred
  reflection models detailed in Table 1.  In each panel, the FPMA
  spectrum is shown in black, and the FPMB spectrum in shown in red.
  TOP LEFT: {\it relconv\_lp\_ext*xillver}, the best overall fit.  TOP
  RIGHT: {\it relxilllp}, a close second-best fit.  BOTTOM LEFT: {\it relxill},
  which does not assume a lamp post geometry.  BOTTOM RIGHT:
  {\it relconv*reflionx}, which makes no geometrical assumption, and
  utilizes a different family of disk reflection models.}
\end{figure}
\medskip

\begin{figure}
\hspace{0.5in}
\includegraphics[scale=0.4]{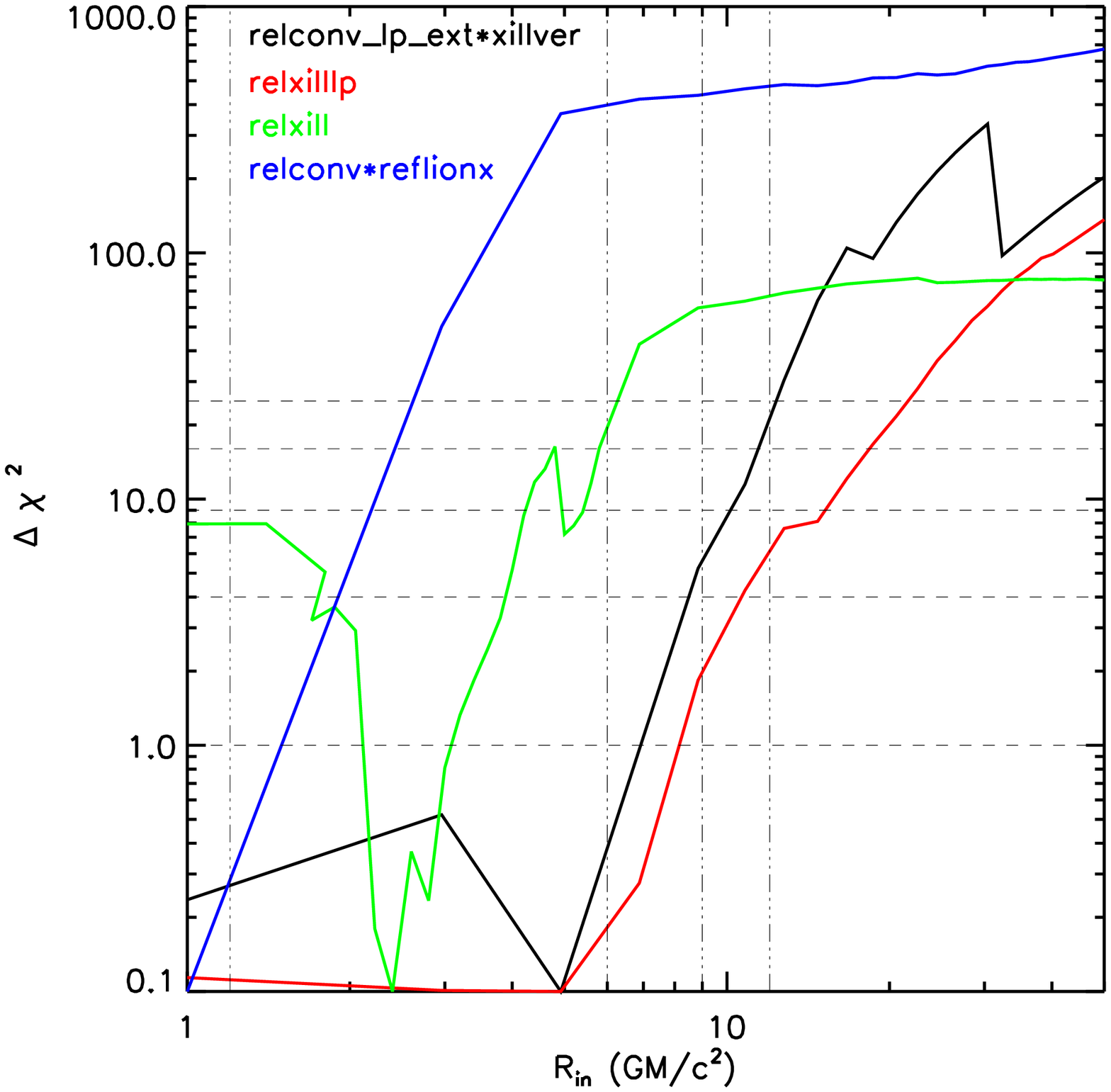}
\includegraphics[scale=0.4]{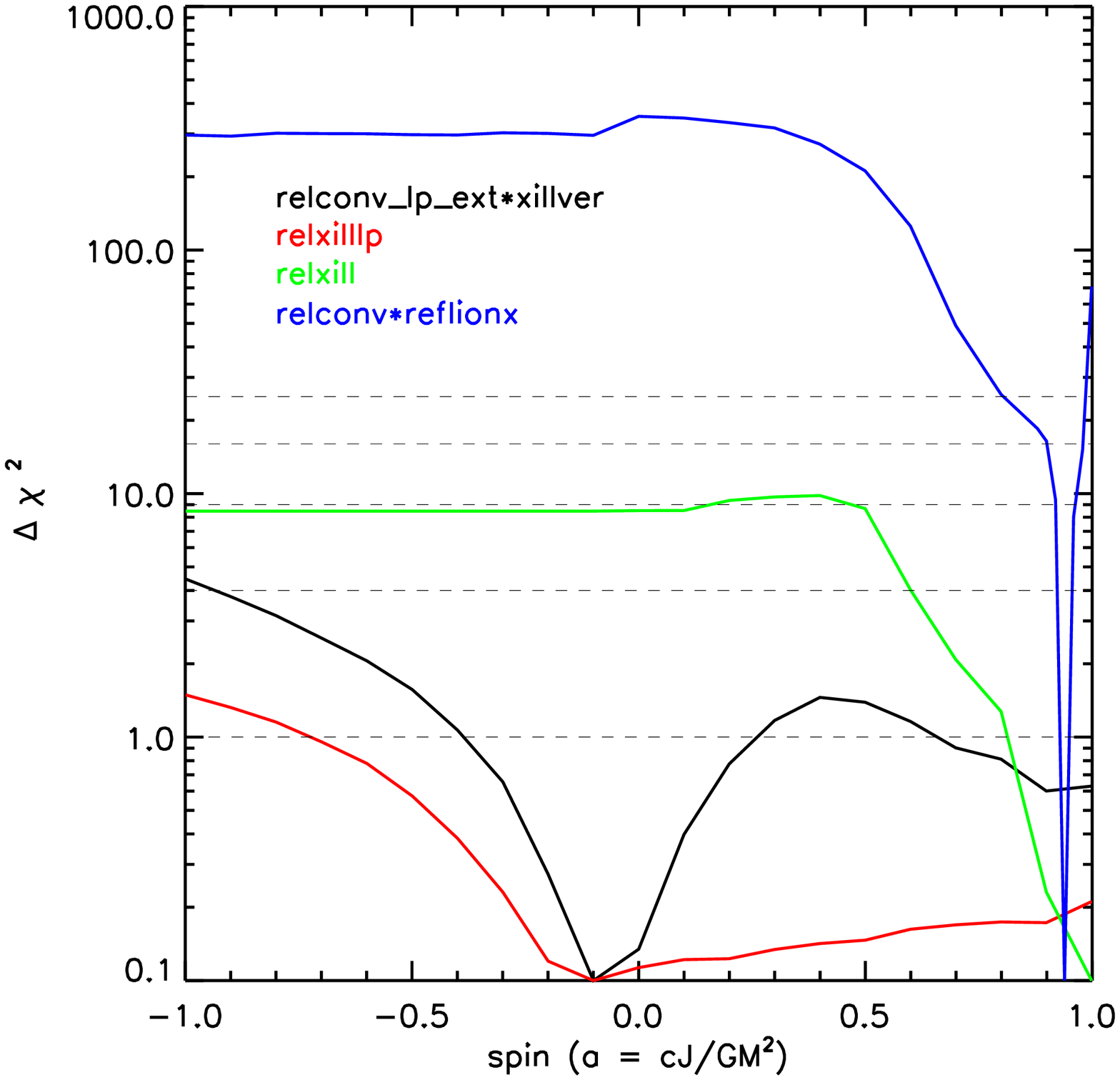}
\figcaption[t]{\footnotesize Confidence levels for key physical
  parameters of the relativistically blurred reflection models in
  Table 1.  LEFT: The change in the goodness-of-fit statistic,
  $\chi^{2}$, is plotted versus the inner disk radius.  The dashed and
  dotted vertical lines indicate the ISCO radius for prograde and
  retrograde accretion disks, for both maximal and zero black hole
  spin.  RIGHT: The change in the goodness-of-fit statistic is plotted
  versus the spin parameter of the black hole.  In both panels, the dashed
  horizontal lines indicate the 1, 2, 3, 4, and $5\sigma$ levels of
  confidence.}
\end{figure}
\medskip

\end{document}